\documentclass[titlepage]{article}
\usepackage{amssymb,epsfig,pslatex}
\usepackage{axodraw}
\usepackage{graphicx}
\usepackage{rotating}
\usepackage{dcolumn}
\usepackage{bm}
\usepackage[latin1]{inputenc}
\usepackage[T1]{fontenc}

\newcommand{\lsim}
{{\;\raise0.3ex\hbox{$<$\kern-0.75em\raise-1.1ex\hbox{$\sim$}}\;}}
\newcommand{\gsim}
{{\;\raise0.3ex\hbox{$>$\kern-0.75em\raise-1.1ex\hbox{$\sim$}}\;}}

\newcommand{\comment}[1]{}
\begin{document}
\begin{titlepage}
\noindent
{\large hep-ph/0605100 \hfill }\\
{\large IISc-CHEP/4/06, LPT-ORSAY-06-32 \hfill \,}
\vspace{0.5cm}
\begin{center}
{\boldmath \huge \bf Lepton distribution as a probe of new physics in 
production and decay of the $t$ quark and its polarization}\\[1.5cm]

{\large \bf Rohini M. Godbole $^{a,1}$, Saurabh D. Rindani $^{b,2}$,
Ritesh K. Singh $^{c,3}$}\\[0.5cm]
$^a$Center for High Energy Physics, IISc, Bangalore, 560 012, INDIA\\
$^b$Physical Research Laboratory, Navarangpura, Ahmedabad 380 009,
INDIA\\
$^c$Laboratoire de Physique Th\'eorique,  91405 Orsay Cedex, FRANCE\\[0.1cm]
$^1${\it rohini@cts.iisc.ernet.in},
$^2${\it saurabh@prl.res.in},
$^3${\it ritesh.singh@th.u-psud.fr}\\[1.0cm]

{\large \bf Abstract}\\[0.2cm]
\parbox[t]{\textwidth}{
We investigate the possibility of studying new physics in various
processes of $t$-quark production using kinematical distributions of the
secondary lepton coming from the decay of $t$ quarks. We show that the angular
distribution of the secondary lepton is insensitive to the anomalous $tbW$
vertex and hence is a pure probe of new physics in a generic process of 
$t$-quark
production. The energy distribution of the lepton is distinctly affected
by anomalous $tbW$ couplings and can be used to analyze them independent of
the production process of $t$ quarks. The effects of $t$ polarization on the
distributions of the decay lepton are demonstrated for top-pair production
process at a $\gamma\gamma$ collider mediated by a heavy Higgs boson.}
\end{center}
\vspace{-0.3cm}
PACS number(s)~: 14.65.Ha, 13.88.+e, 13.85.Qk\\
Keywords~:top, polarization, anomalous top coupling\\
\hrule\vspace{-0.5cm}
\tableofcontents
\vspace{0.3cm}
\hrule
\end{titlepage}
\noindent
\boldmath
\section{Introduction}
The mechanism of spontaneous symmetry breaking (SSB), which is
responsible for generating masses for all fermions and weak
bosons, still lacks explicit experimental verification. In the 
Standard Model (SM), Higgs mechanism is responsible for the SSB and
the Higgs boson, being a remnant degree of freedom after symmetry breaking, 
carries information about the phenomenon of symmetry breaking. The
top quark, whose mass is very close to the electroweak symmetry
breaking scale, is expected to provide a probe for understanding 
SSB in the SM. Direct experimental observation of the Higgs boson is essential 
for establishing the Higgs mechanism as {\it the} correct SSB mechanism. 

The SM has been tested to be {\em the} model of particle
interactions for all the particles other than 
the $t$ quark, which has not yet been studied extensively at the colliders and
the Higgs boson, which is yet to be observed experimentally. 
In addition to the discovery of the Higgs boson, it is also
essential to measure accurately  the couplings of the Higgs boson and the top 
quark to other SM particles with high precision. If these couplings would be
found to be the same as those predicted in the SM then it will confirm the
Higgs mechanism of SSB. Any deviation will signal presence of physics
beyond the SM.

In spite of the impressive agreement of {\it all} the precision electroweak
(EW) measurements with the predictions of the SM, it still suffers from
quite a few deficiencies from a theoretical point of view. For example,
the mass of the
SM Higgs boson is not stable against radiative corrections; also the
SM  does not provide a first principle understanding of the phenomenon of
$CP$ violation, even though it does contain a successful parametrization of the
same in terms of the CKM phase, etc.  All attempts to cure these
and other ills of the SM  require us to go beyond the SM. Such physics beyond 
the SM will imply deviations of the  couplings  of the Higgs boson and the top 
quark with each other as well as with other SM particles.
The specific deviations of the top quark couplings  from the expectations of 
the SM may depend on the details of the particular extension of the SM 
one is looking at.  In this work, we adopt a model-independent formulation and 
allow, in the effective theory approach, the most general interaction of $t$  
with other SM particles.  For example, the most general expression for the 
$tbW$ vertex may be written as,
\begin{eqnarray}
\Gamma^\mu&=&\frac{-i g}{\sqrt{2}} \ \left[\gamma^\mu(f_{1L}P_L + f_{1R}P_R)
- \frac{i\sigma^{\mu\nu}}{m_W} \ (p_t-p_b)_\nu \ (f_{2L}P_L +
f_{2R}P_R) \right].
\label{V:tbW}
\end{eqnarray}
For the SM,  $f_{1L} = 1$ and the anomalous couplings $f_{1R} = f_{2L}
= f_{2R} = 0$. The various extensions of the SM would have specific 
predictions for these anomalous couplings. Since in a renormalizable theory,
these can arise only at a higher order in perturbation theory, we assume them 
to  be small and retain only terms linear in them.
The couplings of the $t$ quark with other gauge bosons can also be
parametrized in a model-independent way similar to Eq.~(\ref{V:tbW}).

These  non-standard couplings  may give rise to changes in the kinematical 
distributions and  polarization of the produced $t$ quarks. Kinematic 
distributions of the decay products of the polarized top quark can yield 
information on its polarization and can be used to construct probes of the 
anomalous top-quark couplings   involved in their production. Such analysis
is simplified if one can devise observables
which are sensitive only to the anomalous coupling
involved in the production process and are independent of a possible 
anomalous $tbW$ vertex. We call these observables  {\it decoupled
observables}. Such observables when used in conjunction with the
remaining observables may be also yield information about the anomalous
$tbW$ vertex itself.  The angular distribution of the secondary lepton coming
from the decay of the $t$ quark is a decoupled observable. 
The energy distribution 
of the secondary lepton, on the other hand, depends upon the anomalous
$tbW$ couplings along with possible new physics in the production
of the $t$ quark. The angular distribution of the $b$ quark from the
decay of the $t$ quark is also sensitive to the $t$ polarization as
well as to the anomalous $tbW$ vertex. Note that the angular distribution
of the decay lepton in the rest frame of the $t$ quark involves only 
the polarization of the parent quark.

The independence of the lepton angular distribution from the anomalous
$tbW$ coupling has been observed for $e^+e^-\to t\bar
t$~\cite{eett1,eett2} and $\gamma \gamma \to t\bar
t$~\cite{phphtt1,phphtt2} earlier, neglecting the
$b$ quark mass. It has been shown that such a result holds independent 
of the initial state~\cite{decoup1} for a massless $b$ quark and also 
applies to any inclusive $t$-quark production process  for a massive 
$b$ quark~\cite{decoup2}.  This possibility gives us a tool to study  
any non-SM physics involved in $t$-quark production at all colliders. 
In this paper we extend our earlier 
analysis~\cite{eett1,phphtt2} to $2\to n$
reactions with more relaxed assumptions on the kinematics and
discuss the use of lepton distributions in reconstructing the
polarization of the $t$ quark in a generic production process.

Polarization of  the $t$ quark is a good probe of new physics
beyond the SM including $CP$ violation. It can be estimated using 
the shape of the distributions of its decay products~\cite{tpol:Dalitz,
tpol:Sehgal, tpol:Grzadkowski2, tpol:Espriu, tpol:Christova} or the 
polarization of the $W$~\cite{tpol:Grzadkowski,tpol:Fischer,tpol:Nelson}.
Recently, a more realistic study of top quark spin measurement
has been performed~\cite{Tsuno:2005qb} using a newly devised
method~\cite{Sumino:2005pg}. Top polarization in the SM has been studied in 
great detail: at tree level in~\cite{Terazawa:1995bd, Atag:2004cp}, including
electroweak corrections in~\cite{Yuan:1991nt, Ladinsky:1992bc},
including QCD corrections in~\cite{Bernreuther:1995cx,Akatsu:1997tq}
and  including electroweak as well as QCD corrections 
in~\cite{Bernreuther:1992ef,Kane:1991bg}. Further, threshold
effects in $t\bar t$ production and top polarization  have been
studied within the framework of the SM in 
Refs.~\cite{Harlander:1994ac, Jezabek:1996re,
Harlander:1996vg, Chibisov:1998vq,Awramik:2001ut,Zhang:2003ph}. 
Spin correlations in top pair production have been 
studied~\cite{Parke:1996pr,Brandenburg:1996df,Mahlon:1997uc,
Hori:1997fe,Brandenburg:2005uu}.  QCD 
corrections to such correlations have also been 
studied~\cite{Tung:1997ur,Brandenburg:1998xw, Kodaira:1998gt,
Bernreuther:2000yn,Kiyo:2000ag,Kiyo:2000th,Bernreuther:2001rq,
Bernreuther:2001bx}.

In this paper we study the polarization of the $t$ quark in a
generic production process, which may receive contributions 
from  new physics, using kinematical distributions of secondary leptons, 
as well as those of $b$ quarks.  We also consider the possibility of 
probing the new physics contribution to top production and  decay  
separately. 

Our main results may be summarized as follows. The lepton angular
distribution is shown to be completely insensitive to any
anomalous $tbW$ coupling assuming a narrow-width approximation for
the $t$ quark and keeping only terms linear in anomalous $tbW$
couplings for any top-production process. 
The decay lepton energy distribution in the rest frame of the $t$ quark,
on the other hand,  is sensitive only to the anomalous $tbW$ couplings.
Specific asymmetries involving lepton angular distribution
relative to the top momentum can  be constructed which measure the top 
polarization in a generic process of $t$-quark production.

The rest of the paper is organized as follows. In Section 2 we
generalize the decoupling theorem for non-zero mass of the $b$ quark
and keeping all the anomalous couplings in Eq.~(\ref{V:tbW})
non-zero. We identify the main ingredients in arriving at the
decoupling theorem and in Section 3 we extend it to a generic
$2\to n$ process of $t$-quark production. We also discuss the effect of
the inclusion  of radiative corrections in our analysis on the validity of 
this decoupling theorem.  In Section 4 we construct lepton angular asymmetries
to reconstruct the polarization density matrix of the decaying $t$
for any generic process. In Section 5 we use the energy
distribution of the secondary lepton to probe anomalous $tbW$
couplings and also discuss the possibility of probing $t$
polarization using $E_\ell$ distributions. In Section 6, we
demonstrate the effects of $t$ polarization on the angular
distribution of the decay leptons from the $t$ quark produced in the
process $\gamma\gamma\to t\bar t$, where the production also includes 
contribution coming from the Higgs-boson mediated diagram.
Section 7 discusses the use of the $b$-quark
angular distribution in conjunction with the lepton angular
distribution as a probe of anomalous $tbW$ vertex. We discuss our
results in Section 8 and conclude.

\section{\boldmath  Angular distribution of secondary leptons in $A \ B\to
t \ \bar t$}
\begin{figure}[t]
\centering
\begin{picture}(175,60)(0,0)
\Text(0,26)[rb]{${\bf A}$}         \Text(10,26)[rb]{${\bf B}$}
\LongArrow(20,30)(50,30)
\Text(60,30)[]{$t$}             \Text(70,30)[]{$\bar t$}
\Line(60,23)(60,15)             \LongArrow(60,15)(95,15)
\Text(105,15)[]{$b$}            \Text(115,15)[l]{$W^+$}
\Line(118,8)(118,0)             \LongArrow(118,0)(155,0)
\Text(165,0)[]{$l^+$}           \Text(175,0)[]{$\nu$}
\Line(70,37)(70,45)             \LongArrow(70,45)(95,45)
\Text(105,45)[]{$\bar b$}       \Text(115,45)[l]{$W^-$}
\Line(118,52)(118,60)           \LongArrow(118,60)(155,60)
\Text(165,60)[]{$l^-$}          \Text(175,60)[]{$\bar\nu$}
\end{picture}
\caption{The  diagram depicting  generic $2\to2$ process of $t$-quark pair
production and subsequent decays. \label{fig:2to2}}
\end{figure}
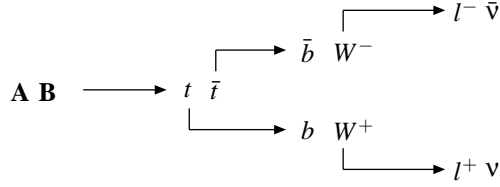

We first look at $t \bar t$ production at either an $e^+e^-$ or a 
$\gamma\gamma$ collider followed by the decay of $t/\bar t$  into 
secondary leptons.  We take the most general $tbW$ vertex. The  process 
is shown in Fig~\ref{fig:2to2}.
The square of the matrix element for this process, including semi-leptonic
decay of  $t$ and inclusive decay of $\bar t$, can
be written using the narrow width approximation for the $t$ quark as
\begin{eqnarray}
\overline{|{\cal M}|^2} = \frac{\pi \delta(p_t^2-m_t^2)}{\Gamma_t m_t}
\sum_{\lambda,\lambda'} \rho(\lambda,\lambda')\Gamma(\lambda,\lambda').
\end{eqnarray}
Here we have
\begin{eqnarray}
\rho(\lambda,\lambda') = M_\rho(\lambda) \ M^*_\rho(\lambda')
\hspace{0.5cm} \mbox{and}\hspace{0.5cm}
\Gamma(\lambda,\lambda') = M_\Gamma(\lambda) \ M^*_\Gamma(\lambda'),
\end{eqnarray}
where $M_\rho(\lambda)$ is the production amplitude of a $t$ quark
with helicity $\lambda$, and $M_\Gamma(\lambda)$ is the decay
matrix element of a $t$ quark with helicity $\lambda$. 
We study  the process $A+B \to \bar t b \nu \ell^+$ where the $\bar t$ decays
inclusively. The differential cross-section for this process can be written as
\begin{eqnarray}
d\sigma = \frac{(2\pi)^4}{2I} \ \overline{|{\cal M}|^2} \
\delta^4(k_A+k_B-p_{\bar t}-p_b-p_\nu-p_\ell) \ \frac{d^3p_{\bar
t}}{2E_{\bar t}(2\pi)^3} \ \frac{d^3p_{b}}{2E_{b}(2\pi)^3} \
\frac{d^3p_{\nu}}{2E_{\nu}(2\pi)^3} \
\frac{d^3p_{\ell}}{2E_{\ell}(2\pi)^3}, \label{dsig22}
\end{eqnarray}
where $I^2=[s-(m_A+m_B)^2][s-(m_A-m_B)^2] $.
Using the expression for $ \overline{|{\cal M}|^2}$ and inserting
\begin{eqnarray}
1 = \intop d^4p_t \delta^4(p_t-p_b-p_\nu-p_\ell)
  = \intop d(p_{0}^2) \frac{d^3p_t}{2p_0} \ \theta(p_0^2) \ \delta^4
  (p_t-p_b-p_\nu-p_\ell)
\end{eqnarray}
in Eq.~(\ref{dsig22}), we can rewrite the differential cross-section after
integrating over $p_0^2$ as 
\begin{eqnarray}
d\sigma &=& \sum_{\lambda,\lambda'} \left[
\frac{(2\pi)^4}{2I} \rho(\lambda,\lambda')\delta^4(k_A+k_B-p_{\bar t}-p_t)
\frac{d^3p_{\bar t}}{2E_{\bar t}(2\pi)^3} \ \frac{d^3p_{t}}{2E_{t}(2\pi)^3}
\right]\nonumber \\
&\times&\left[\frac{1}{\Gamma_t} \ \left(
\frac{(2\pi)^4}{2m_t} \Gamma(\lambda,\lambda')\delta^4(p_t-p_b-p_\nu-p_\ell)
\frac{d^3p_{b}}{2E_{b}(2\pi)^3} \ \frac{d^3p_{\nu}}{2E_{\nu}(2\pi)^3} \right)
\frac{d^3p_{\ell}}{2E_{\ell}(2\pi)^3} \right].
\label{facsig}
\end{eqnarray}
The narrow-width approximation for the $t$ quark plays a  crucial role 
in arriving at Eq.~(\ref{facsig}), where the
differential cross-section for the full process is expressed as the
product of the differential cross-section for $t\bar t$ production
and the differential decay rate of the $t$ quark. The term in the
first pair of square brackets in Eq.~(\ref{facsig}) can be written as
\begin{eqnarray}
\intop
\frac{d^3p_{\bar t}}{2E_{\bar t}(2\pi)^3} \ \frac{d^3p_{t}}{2E_{t}(2\pi)^3}
\ \frac{(2\pi)^4}{2I} \ \rho(\lambda,\lambda') \ \delta^4(k_A+k_B-p_{\bar t}
-p_t) = d\sigma_{2\to2}(\lambda,\lambda') d\cos\theta_t \ .
\end{eqnarray}
Similarly, the term in the second square bracket in
Eq.~(\ref{facsig}) can be integrated in the rest frame of the $t$
quark to give
\begin{eqnarray}
&&\frac{1}{\Gamma_t} \ \frac{(2\pi)^4}{2m_t} \intop
\frac{d^3p_{\ell}}{2E_{\ell}(2\pi)^3} \ \
\frac{d^3p_{b}}{2E_{b}(2\pi)^3} \ \frac{d^3p_{\nu}}{2E_{\nu}(2\pi)^3} \ \
\Gamma(\lambda,\lambda')\delta^4(p_t-p_b-p_\nu-p_\ell) \nonumber\\
&=&\frac{1}{\Gamma_t} \ \frac{(2\pi)^4}{2m_t} \intop
\frac{d^3p_{\ell}}{2E_{\ell}(2\pi)^3} \ \ dE_b \ d\phi_b \ \
\frac{1}{4 \ E_\ell}  \frac{1}{(2\pi)^6}
\Gamma(\lambda,\lambda')\nonumber\\
&=& \frac{1}{32\Gamma_t m_t} \ \frac{E_\ell}{(2\pi)^4} \frac{\langle
\Gamma(\lambda,\lambda')\rangle}{m_t E_\ell} \ \ dE_\ell \
d\Omega_\ell \ dp_W^2 \ .
\end{eqnarray}
Here angular brackets denote an average over the azimuthal angle
of the $b$ quark w.r.t the  plane of the $t$ and the $\ell$ momenta
chosen as the $x-z$ plane, where the $z$ axis points
in the direction of the lepton momentum. We first change the
angular variables of the $b$ quark from $[\cos\theta_b, \ \phi_b]$
to $[\cos\theta_{b\ell}, \ \phi_{b\ell}]$ and then average over
$\phi_{b\ell}$. Further, the integral over $E_b$ is replaced by an
integral over the invariant mass of the $W$ boson, $p_W^2$.
Boosting the above expressions to the c.m. frame one can rewrite
Eq.(\ref{facsig}) as
\begin{eqnarray}
d\sigma = \frac{1}{32 \ \Gamma_t m_t} \ \frac{E_\ell}
{(2\pi)^4} \left[ \sum_{\lambda,\lambda'} d\sigma_{2\to2}(\lambda,\lambda') \
\times \ \left(\frac{\langle\Gamma'(\lambda,\lambda')\rangle}{m_t E_\ell^0}
\right)_{\rm c.m.} \right] \
d\cos\theta_t \ dE_\ell \ d\cos\theta_\ell \ d\phi_\ell \ dp_W^2,
\end{eqnarray}
where $E^0_\ell$ is the lepton energy in the rest frame of the $t$
quark.

In the rest frame of the $t$ quark with the $z$ axis along the
direction of the boost of the $t$ quark to the lab frame, and the $x-z$ plane
coincident with the $x-z$ plane of the lab frame, the expressions
for $\langle\Gamma(\lambda,\lambda')\rangle$ are given by
\begin{eqnarray}
\langle\Gamma(\pm,\pm) \rangle
&=& g^4 m_t E_\ell^0 |\Delta_W(p_W^2)|^2 \ (1\pm\cos\theta_l) \
\times \ F(E_\ell^0), \nonumber\\
\langle\Gamma(\pm,\mp) \rangle
&=& g^4 m_t E_\ell^0 |\Delta_W(p_W^2)|^2 \ (\sin\theta_le^{\pm
i\phi_l}) \ \times \ F(E_\ell^0).
\label{ddmat}
\end{eqnarray}
Here $\Delta(p_W^2)$ is the $W$-boson propagator and
$F(E_\ell^0)$ is given by
\begin{eqnarray}
F(E_\ell^0) &=&
\left[(m_t^2-m_b^2-2p_t\cdot p_l)\left(|f_{1L}|^2 + \Re(f_{1L}f_{2R}^*)
\frac{m_t}{m_W} \frac{p_W^2}{p_t.p_l} \right) \right.\nonumber\\
&-& \left.2\Re(f_{1L}f_{2L}^*) \ \frac{m_b}{m_W} \ p_W^2 - \Re(f_{1L}f_{1R}^*)
\ \frac{m_b \ m_t}{p_t.p_l} \ p_W^2 \right].
\end{eqnarray}
Eq.~(\ref{ddmat}) assumes that all the anomalous $tbW$ couplings other than
$f_{1L}$ are small, and terms quadratic in them are dropped.
The azimuthal correlation between $b$
and $\ell$ is sensitive to the anomalous $tbW$ couplings. The averaging
eliminates any such dependence and we get
$\langle\Gamma(\lambda,\lambda')\rangle$ factorized
into angular part $A(\lambda,\lambda')$ and energy dependent part
$F(E_\ell^0)$.
In short the expression for decay density matrix can be written as
\begin{eqnarray}
\langle\Gamma(\lambda,\lambda')\rangle= (m_t E_\ell^0) \ |\Delta(p_W^2)|^2 \
g^4 A(\lambda,\lambda') \  F(E_\ell^0).
\end{eqnarray}
Here $A(\lambda,\lambda')$ depends only on the polar and azimuthal angles 
of $\ell$ in the rest frame of $t$ and $F(E_l^0)$ depends only on the lepton
energy, various masses and couplings. After boosting to the c.m. or lab frame,
they pick up additional dependence on $E_t$ and $\theta_t$. The most important
point is that $\langle\Gamma(\lambda, \lambda')\rangle$
factorizes into a pure angular part, $A(\lambda,\lambda')$, and a pure energy 
dependent part, $F(E_\ell)$. Thus the angular
dependence of the density matrix remains insensitive to the anomalous $tbW$
couplings up to an overall factor $F(E_\ell)$. Putting the expression
for the
decay density matrix in Eq.~(\ref{facsig}) we get
\begin{eqnarray}
d\sigma&=&\frac{1}{32 \ \Gamma_t m_t} \ \frac{1}
{(2\pi)^4} \left[ \sum_{\lambda,\lambda'} d\sigma_{2\to2}(\lambda,\lambda') \
\times \ g^4 A(\lambda,\lambda') \right] \
d\cos\theta_t \ d\cos\theta_\ell \ d\phi_\ell\nonumber \\
&\times& E_\ell \ F(E_\ell) \ dE_\ell \ \ dp_W^2.
\label{dsigell}
\end{eqnarray}
Since the $E_\ell$-dependent part has factored out, one can integrate 
this out. The limits of integration for $E_\ell$ in the c.m. frame are
given by
$$ \frac{p_W^2}{2E_t} \frac{1}{1-\beta_t\cos\theta_{t\ell}} \le E_\ell \le
\frac{m_t^2-m_b^2}{2E_t} \frac{1}{1-\beta_t\cos\theta_{t\ell}} \ ,$$
and after integration we get
\begin{eqnarray}
&&\int dE_\ell \ E_\ell F(E_\ell) = \frac{1}{E_t^2 \ (1-\beta_t \
\cos\theta_{t\ell})^2} \Bigg[ -\frac{|f_{1L}|^2}{12} \ \bigg( (m_t^2-m_b^2)^3 -
(p_W^2)^3 \bigg)\nonumber \\
&&+\Bigg( |f_{1L}|^2(m_t^2-m_b^2) - 2\Re(f_{1L}f_{2R}^*)\frac{m_t p_W^2}{m_W}
-2\Re(f_{1L}f_{2L}^*)\frac{m_b p_W^2}{m_W}\Bigg)\frac{(m_t^2-m_b^2)^2 -
(p_W^2)^2 }{8}\nonumber \\
&&+\Bigg((m_t^2-m_b^2) \Re(f_{1L}f_{2R}^*)\frac{m_tp_W^2}{m_W} -
\Re(f_{1L}f_{1R}^*) m_b m_t p_W^2\Bigg)(m_t^2-m_b^2-p_W^2) \Bigg]\\
&&=\frac{G(\{m_i\},\{f_i\},p_W^2)}{E_t^2 \ (1 - \beta_t \cos\theta_{t\ell})^2}.
\label{gfac}
\end{eqnarray}
Here $G$ depends upon masses $\{m_i\}$, couplings $\{f_i\}$ and 
$p_W^2$. The same factor $G$ appears in the expression for the decay width
$\Gamma_t$ as well, and cancels in Eq. (\ref{dsigell}) after integration over
$p_W^2$, leaving the differential rate independent of
any anomalous $tbW$ vertex. This {\it decoupling} of lepton distribution from
the anomalous $tbW$ couplings has been shown using the same method in
Ref.~\cite{phphtt2} for massless $b$ quarks for the case of $\gamma\gamma \to t
\bar t$. Here we  extend the decoupling result to include (1) a
massive $b$ quark, (2) all the anomalous $tbW$ couplings and (3) finite width
of the $W$ boson. The important approximations/assumptions
in arriving at this result are~:
\begin{enumerate}
\item Narrow-width approximation for the $t$ quark.
\item Smallness of $f_{1R}, \ f_{2L}$ and $f_{2R}$.
\item $t \to bW (\nu\ell)$ as the only decay channel of the $t$ quark.
\end{enumerate}
The first of these, viz., the narrow-width approximation for the $t$ quark, 
is used to factorize the differential cross-section into the production and 
the decay of the $t$ quark as shown in Eq.~(\ref{facsig}). The effect of the 
finite-width corrections on normalized distributions of the decay products is 
expected to be negligible. An example of explicit verification of the fact can
be found in Ref.~\cite{Aguilar-Saavedra:2006fy}.  The second assumption, i.e. 
the smallness of the anomalous couplings is essential for the factorization of
$\langle\Gamma(\lambda,\lambda')\rangle$  into a purely angular and a purely 
energy-dependent part. 
If the lepton spectrum is calculated keeping the quadratic
terms, as would be necessary for large couplings, no factorization of
$\langle\Gamma(\lambda,\lambda')\rangle$ is observed~\cite{Najafabadi}.
\comment
{
For large anomalous coupling, keeping quadratic terms, the
lepton spectrum is calculated in Ref.~\cite{Najafabadi} and no factorization of
$\langle\Gamma(\lambda,\lambda')\rangle$ is observed.}

The third assumption is necessary for {\it exact} cancellation of $G$
in the numerator and $\Gamma_t$. If there are other decay modes of the $t$ quark
than $bW$, then it will result in an extra factor of branching ratio, which
is an overall constant depending upon anomalous coupling. This still maintains
the decoupling of angular distribution of leptons up to an overall scale.
We see that the first two assumptions are the only ones essential to achieving 
the decoupling while the last one only simplifies the calculation. After
factorization of the differential rates into production and decay parts,
the most important
ingredient in achieving decoupling is the factorization of
$\langle\Gamma(\lambda,\lambda')\rangle$ into a purely angular part 
and a purely energy-dependent part, with the angular
part being independent of any anomalous $tbW$ coupling. This factorization is
achieved by averaging over the azimuthal angle of $b$ quarks and keeping only
the linear terms in the anomalous $tbW$ couplings. Thus, as long as the 
anomalous
$tbW$ couplings are small and we do not look for any correlation between
azimuthal angles of $\ell$ and $b$, the lepton angular distributions
remain insensitive to (or decoupled from) any anomalous $tbW$ couplings.

\section{\boldmath  Angular distribution of secondary leptons in $A \ B\to t \ P_1 \ ...
\ P_{n-1}$}
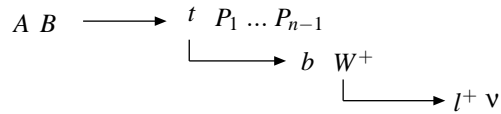
\begin{figure}[ht]
\centering
\begin{picture}(175,40)(0,0)
\Text(0,26)[rb]{$A$}         \Text(10,26)[rb]{$B$}
\LongArrow(20,30)(50,30)
\Text(60,30)[lb]{$t$}             \Text(70,26)[lb]{$P_1 \ ... \ P_{n-1}$}
\Line(60,23)(60,15)             \LongArrow(60,15)(95,15)
\Text(105,15)[]{$b$}            \Text(115,15)[l]{$W^+$}
\Line(118,8)(118,0)             \LongArrow(118,0)(155,0)
\Text(165,0)[]{$l^+$}           \Text(175,0)[]{$\nu$}
\end{picture}
\caption{\label{fig:2ton} Diagram depicting  $2\to n$ process}
\end{figure}

After identifying the requisites to arriving at the decoupling of the lepton
distribution from an anomalous $tbW$ vertex, we intend to look at 
the  production of the $t$ quark in a generic process
$A B \rightarrow t P_1 P_2 ..P_n$, followed by its semi-leptonic decay. 
In this section we assume only the narrow-width approximation for $t$ quarks 
and smallness of the anomalous $tbW$ couplings. 
A representative diagram of the  process is shown in Fig.~\ref{fig:2ton}.
The final state particles $P_i$ may decay inclusively. After using the
narrow-width approximation for the $t$ quark, the expression for the 
differential
cross-section, similar to Eq.~(\ref{facsig}), can be written as~:
\begin{eqnarray}
d\sigma &=& \sum_{\lambda,\lambda'} \left[
\frac{(2\pi)^4}{2I} \rho(\lambda,\lambda')
\delta^4\bigg(k_A+k_B-p_t-\bigg(\sum_i^{n-1}p_i\bigg) \bigg)
\frac{d^3p_{t}}{2E_{t}(2\pi)^3} \ \prod_i^{n-1}
\frac{d^3p_{i}}{2E_{i}(2\pi)^3} \right]\nonumber \\
&\times&\left[\frac{1}{\Gamma_t} \
\frac{(2\pi)^4}{2m_t} \Gamma(\lambda,\lambda')\delta^4(p_t-p_b-p_\nu-p_\ell)
\frac{d^3p_{b}}{2E_{b}(2\pi)^3} \ \frac{d^3p_{\nu}}{2E_{\nu}(2\pi)^3}
\frac{d^3p_{\ell}}{2E_{\ell}(2\pi)^3} \right].
\label{facsign}
\end{eqnarray}
In the c.m. frame we choose a set of axes such that the production plane of the 
$t$ quark defines the
azimuthal reference $\phi=0$ and rewrite the production part as~
\begin{eqnarray}
\intop
\frac{d^3p_{t}}{2E_{t}(2\pi)^3} \ \prod_i^{n-1}
\frac{d^3p_{i}}{2E_{i}(2\pi)^3} \ \
\frac{(2\pi)^4}{2I} \rho(\lambda,\lambda')
\delta^4\bigg(k_A+k_B-p_t-\bigg(\sum_i^{n-1}p_i\bigg) \bigg)
= d\sigma_{2\to n}(\lambda,\lambda') \ dE_t
d\cos\theta_t.
\end{eqnarray}
We will continue to use the symbol
$d\sigma_{2\to n}(\lambda,\lambda')$ independent of whether the above integral 
can be done in a closed form or has to be done numerically.  Using this we 
write the expression for $d\sigma$ similar to Eq.~(\ref{dsigell}) as~
\begin{eqnarray}
d\sigma &=& \frac{1}{32 \ \Gamma_t m_t (2\pi)^4}
\left[ \sum_{\lambda,\lambda'} d\sigma_{2\to n}(\lambda,\lambda') \
\times \ g^4A(\lambda,\lambda') \right] \
dE_t \ d\cos\theta_t \ d\cos\theta_\ell \ d\phi_\ell \nonumber\\
&\times& E_\ell \ F(E_\ell) \  dE_\ell \ dp_W^2.
\label{dsig2n}
\end{eqnarray}
Again, the $E_\ell$ integration will give the same factor as that appearing in
the expression for $\Gamma_t$ which will cancel between the numerator
and denominator as in the case of $2\to2$ process
of $t$-quark production. Thus we have demonstrated the decoupling of angular
distribution of the leptons from anomalous $tbW$ couplings for a most
general $2\to n$ production process for the $t$ quark. 
Any observable constructed
using the angular variables of the secondary lepton will thus be completely
independent of the anomalous $tbW$ vertex. Hence it is a pure probe of couplings
involved in the corresponding production process and the effect of any anomalous
coupling in decay process has been {\em filtered} out by averaging over
the
azimuthal angle of the $b$ quark and the energy of the decay leptons. 
Further, for
hadronic decay of the $t$ quark $W\to u\bar d$, where $u$ stands for $u$
and $c$ quarks, and
$d$ stands for $d$ and $s$ quarks, the decoupling goes through. 
The role of $\ell$ is taken by down-type
quarks, the $T_3=-1/2$ fermions in the $SU(2)$ doublet. For construction of
these angular distributions one needs to distinguish the $\bar d$-jet
from the $u$-jet, which requires charge determination of light quarks. This,
unfortunately, is not possible. Thus, though $\bar d$-jet provides 
a high-statistics {\em decoupled} angular distribution, it cannot be used and 
in real experiments we have access only to the lepton  angular distribution.

The angular distribution of leptons has been used to probe new physics in
various processes of $t\bar t$ production at a Linear Collider~\cite{LC:lept} 
and a photon collider~\cite{PLC:lept}. All of these studies had assumed a 
massless 
$b$ quark and hence included the effect  only of $f_{2R}$ from  the anomalous
part of the $tbW$ vertex. With the above proof of decoupling, now all the 
results of Ref.~\cite{LC:lept,PLC:lept} can be extended to include the case of
a massive $b$ quark, and thus the case of all the anomalous $tbW$ couplings
being nonzero.

This decoupling has been demonstrated only after 
averaging over the entire allowed range of the lepton energy. One might
worry that  
an experiment will always involve a lower cut on the energy
$E_\ell^{lab}$ of the 
lepton in the laboratory frame  which may be higher than the minimum
$E_\ell^{lab}$ allowed kinematically. 
In fact for $t$ quarks with velocity $\sim 0.7 \ c$ in 
the laboratory  frame, the minimum value of $E_\ell^{lab}$ is about 8 GeV and 
thus a cut will cause no problem for the validity of the decoupling.
In our numerical analysis for the chosen collider energy,
the  minimum energy of the lepton in the laboratory 
is always above the typical energy cut.

{\bf Radiative corrections~:} We now examine to what extent the 
general proof of 
the decoupling of the lepton angular distribution for $2\to n$ process goes 
through even after including the radiative corrections. 
Radiative correction to the full process involves correction to
the production process alone, correction to the decay process alone and the
non-factorizable correction. We consider these one by one.

Radiative corrections to the $2\to n$ production process include a  
$2\to n$ process with a loop and a  $2\to n+1$ process for a real photon 
(gluon) emission. As the factorization outlined in Eq. (\ref{dsig2n}) is
independent of the number $n$ of particles in the final state, it takes
place in case of both the above corrections.  
Thus, radiative corrections to the production process do not in any way
modify the independence of the decay-lepton angular distribution from
anomalous interactions in the decay vertex.
It may be noted that these radiative corrections
do of course modify the 
$\rho(\lambda,\lambda')$ representing the production density matrix.

Virtual photon (gluon) correction to the decay process can be
parametrized in terms of various anomalous $tbW$ couplings shown in
Eq.~(\ref{V:tbW}). However real photon (gluon) emission from decay products
of the $t$ quark can alter the energy and angle factorization 
of $\Gamma(\lambda, \lambda')$. For semi-leptonic decay, the QCD correction 
to the above factorization is very small, at the per-mill level~\cite{lepQCD}. 
\comment{
This conclusion does not assume soft gluons. Nevertheless, since we concentrate only on corrections arising from soft emissions, it is safe to assume that the 
result is true for soft gluons as well.  
}
Thus, the accuracy of factorization on 
inclusion of radiative corrections involving real gluon emission would be 
at the per-mill level, and would remain similar  on inclusion of 
real photon emission as well.

For hadronic decays of the $t$ quark, i.e. for $W^+\to u\bar d (c\bar s)$, the
angular distribution of $\bar d$ receives additional QCD correction as compared
to the leptonic channel. These QCD corrections to $t\to b u \bar d$ are large,
about 7\%~\cite{Brandenburg:2002xr}.  The factorization of 
$\Gamma(\lambda,\lambda')$ receives a large correction.  The $\bar d$ angular 
distribution is therefore modified substantially due to radiative corrections. 
This further favors the use  of semi-leptonic decay channel for the 
$t$ quark in the analysis of various new physics issues.

We now address the issue of non-factorizable corrections.
These have been calculated for different $2\to2$
processes for $t\bar t$ production and subsequent
decays~\cite{nfQCD1,nfQCD2,Kolodziej:2005rx,nfQCD3,nfQCD4} and  depend
on the particular  kinematic variable being considered.   
For the invariant-mass distribution 
of the virtual quark, for example, they could be as large as 100\% for an 
$e^+e^-$ initial state near $t\bar t$ threshold. However, the magnitude of 
these corrections gets smaller as one goes away from the threshold. 
The enhancement near threshold is caused by increased importance of 
Coulomb type interaction between the slowly moving decay products.  
Most importantly, either near the threshold or far above it, the correction 
is exactly zero when the  $t$ quark is on shell~\cite{nfQCD1} and the 
corrections  vanish  in the double pole approximation (DPA)
when integrated over the invariant mass of the top decay 
products around the top mass pole~\cite{nfQCD1,nfQCD3,nfQCD4}. 
It thus seems reasonable to extrapolate that  for the energy and
angular distributions of the top decay products, the non-factorizable
corrections would be negligible, if not strictly zero in the
on-shell approximation for $t$ quarks which we use.
This would, however, need to be verified by an actual calculation.

Thus it is very likely that the decoupling of lepton distribution will be valid 
to a good accuracy for radiatively corrected distributions. If this is 
confirmed by explicit calculation of the non-factorizable corrections, 
lepton angular distribution can serve as a robust probe of possible new 
physics in the production process of $t$ quarks. 

\comment{
We must,
however, note that such a decoupling is possible after averaging over the
lepton energy in the entire allowed range. In the lab frame, due to boost
of the $t$ quark, $E_\ell^{lab}$ may be smaller than the usual energy cut.
These soft leptons, with energy below the experimental cut-off may modify the 
decoupling and a weak dependence on anomalous $tbW$-vertex can re-enter. For 
$t$ quarks with velocity $\sim 0.7 \ c$ in the lab frame,  the minimum value 
of $E_\ell^{lab}$ is about 8 GeV, hence can survive the energy cut.
For numerical analysis we choose the collider energy such that the 
minimum energy of the lepton is above the typical energy cut.
}
\section{Secondary lepton distribution and the top-quark polarization}
In this section we will explore probing the $t$-quark polarization through 
lepton angular distributions. We start with Eq.~(\ref{facsign}). The
terms in square brackets are Lorentz invariant by themselves and one can
calculate them in any frame of reference.
Thus we integrate completely the first square
bracket in the rest frame of the top quark, and denote it as
\begin{eqnarray}
\sigma(\lambda,\lambda') =
\intop \frac{d^3p_{t}}{2E_{t}(2\pi)^3} \left(\prod_{i=1}^{n-1}
\frac{d^3p_{i}}{2E_{i}(2\pi)^3}\right) \ \
\frac{(2\pi)^4}{2I} \rho(\lambda,\lambda') \ \ \delta^4\left(k_A+k_B-p_t-
\left(\sum_{i=1}^{n-1} p_i\right)\right).
\label{sigpol}
\end{eqnarray}
Here the total cross-section for our $2\to n$ process  is given by
$\sigma_{tot} = \sigma(+,+) + \sigma(-,-)$, whereas the off-diagonal terms in
$\sigma(\lambda,\lambda')$ are production rates of the $t$ quark with {\em
transverse polarization}. The most general polarization density matrix
of a fermion is parametrized as
\begin{eqnarray}
P_t = \frac{1}{2}\left(
\begin{tabular}{cc}
$1+\eta_3$ & $\eta_1 - i\eta_2$\\
$\eta_1 + i\eta_2$ & $1-\eta_3$
\end{tabular} \right),
\label{poldm}
\end{eqnarray}
where
\begin{eqnarray}
\eta_3 = \left(\sigma(+,+)-\sigma(-,-)\right)/\sigma_{tot} \nonumber\\
\eta_1 = \left(\sigma(+,-)+\sigma(-,+)\right)/\sigma_{tot}\nonumber \\
i \ \eta_2 = \left(\sigma(+,-)-\sigma(-,+)\right)/\sigma_{tot}
\label{polAB}
\end{eqnarray}
and
\begin{equation}
\sigma(\lambda,\lambda')=\sigma_{tot} \ P_t(\lambda,\lambda') \ .
\label{siglL}
\end{equation}
Here $\eta_3$ is the average helicity or the longitudinal polarization and
$\eta_{1,2}$ are two transverse polarizations of the top quark.
Further, using the factorization of
$\langle\Gamma(\lambda,\lambda')\rangle$ into angular and
energy-dependent parts, we can write the second square bracket in
the rest frame of the top quark as
\begin{eqnarray}
\frac{1}{\Gamma_t} \ \frac{(2\pi)^4}{2m_t} \intop
\frac{d^3p_{b}}{2E_{b}(2\pi)^3} \ \frac{d^3p_{\nu}}{2E_{\nu}(2\pi)^3}
\frac{d^3p_{\ell}}{2E_{\ell}(2\pi)^3} \ \ \
\Gamma(\lambda,\lambda')\delta^4(p_t-p_b-p_\nu-p_\ell)
= C \ A(\lambda,\lambda')  \ d\Omega_l.
\label{decaydm}
\end{eqnarray}
In the above, $C$ is a constant obtained after doing all the
integrations other than over $\Omega$,  
and  after cancelling the factor of $G$ in Eq.~(\ref{gfac}) between numerator
and the $\Gamma_t$ in the denominator. Here $A(\lambda,\lambda')$ is the only
factor that depends on the angles of the lepton and is given by
\begin{equation}
A(\pm,\pm) = (1\pm\cos\theta_l), \hspace{1.0cm} A(\pm,\mp) = \sin\theta_l
e^{\pm i \phi_l}
\end{equation}
Using Eqs.~(\ref{sigpol}), (\ref{siglL}) and (\ref{decaydm}),
Eq.~(\ref{facsign}) gives the differential cross-section as
\begin{eqnarray}
\frac{d\sigma}{d\cos\theta_l \ d\phi_l} = C \ \sigma_{tot} \ \left[
1 + \eta_3 \ \cos\theta_l + \eta_1 \ \sin\theta_l\cos\phi_l
+ \eta_2 \ \sin\theta_l\sin\phi_l \right].
\label{dsigrest}
\end{eqnarray}
From the above expression, it is simple to calculate the polarization of the
top quark in terms of polar and azimuthal asymmetries. The expressions for
$\eta_i$ are as follows.
\begin{eqnarray}
\frac{\eta_3}{2}&=&\frac{1}{4\pi \ C \ \sigma_{tot}}
\left[\intop_0^1 d\cos\theta_l \intop_0^{2\pi}d\phi_l
\frac{d\sigma}{d\cos\theta_l \ d\phi_l} - \intop_{-1}^0 d\cos\theta_l
\intop_0^{2\pi}d\phi_l\frac{d\sigma}{d\cos\theta_l \ d\phi_l}\right],\nonumber\\
\frac{\eta_2}{2}&=&\frac{1}{4\pi \ C \ \sigma_{tot}}
\left[\intop_{-1}^1 d\cos\theta_l \intop_0^{\pi}d\phi_l
\frac{d\sigma}{d\cos\theta_l \ d\phi_l} - \intop_{-1}^1 d\cos\theta_l
\intop_{\pi}^{2\pi}d\phi_l \frac{d\sigma}{d\cos\theta_l \ d\phi_l} \right],
\nonumber\\
\frac{\eta_1}{2}&=&\frac{1}{4\pi \ C \ \sigma_{tot}}
\left[\intop_{-1}^1 d\cos\theta_l \intop_{-\pi/2}^{\pi/2}d\phi_l
\frac{d\sigma}{d\cos\theta_l \ d\phi_l} - \intop_{-1}^1 d\cos\theta_l
\intop_{\pi/2}^{3\pi/2}d\phi_l \frac{d\sigma}{d\cos\theta_l \ d\phi_l} \right].
\label{polrest}
\end{eqnarray}
Above, we have $4\pi C = BR(t\to b l \nu)$. These asymmetries can also be
represented in terms of the spin-basis vectors of the $t$ quark. The spin-basis
vectors of the top quark in the rest frame are given by
\begin{eqnarray}
s_1^\mu = (0,1,0,0), \ s_2^\mu = (0,0,1,0), \ s_3^\mu = (0,0,0,1).
\end{eqnarray}
The expressions for the top polarizations given in Eq.~(\ref{polrest}) can also
be written as the following asymmetries~:
\begin{eqnarray}
\frac{\eta_3}{2}&=&\frac{\sigma(p_\ell.s_3 < 0) - \sigma(p_\ell.s_3 > 0)}
{4\pi \ C \ \sigma_{tot}},\nonumber\\
\frac{\eta_2}{2}&=&\frac{\sigma(p_\ell.s_2 < 0) - \sigma(p_\ell.s_2 > 0)}
{4\pi \ C \ \sigma_{tot}},\nonumber\\
\frac{\eta_1}{2}&=&\frac{\sigma(p_\ell.s_1 < 0) - \sigma(p_\ell.s_1 > 0)}
{4\pi \ C \ \sigma_{tot}}.
\label{polgen}
\end{eqnarray}
In other words, the average polarizations of the $t$ quark can be written as
expectation values of the signs of $(p_\ell.s_i)$.
Here we note that Eq.~(\ref{polgen}) is valid in any frame of reference and is
identical to  Eq.~(\ref{polrest}) when $p_\ell.s_i$ are written in the rest
frame of the top quark. We also note that expectation values of 
$(p_\ell.s_i)$ have
been considered in Ref.~\cite{Harlander:1996vg} as probes of $t$-quark
polarization. Our Eq.~(\ref{polgen}), however, relates the average
$t$ polarizations to simple asymmetries, which involves only number-counting
experiments subjected to specific kinematical cuts.

In the lab frame where the top-quark production plane
defines the $\phi_l=0$ plane and its momentum is given by,
$$p_t^\mu = E_t(1,\beta_t\sin\theta_t,0,\beta_t\cos\theta_t),$$ we have
\begin{eqnarray}
s_1^\mu = (0,-\cos\theta_t,0,\sin\theta_t), \
s_2^\mu = (0,0,1,0), \
s_3^\mu = E_t(\beta_t,\sin\theta_t,0,\cos\theta_t)/m_t.
\end{eqnarray}
With this choice of reference frame it is easy to see that $\eta_2$ is the
{\em up-down} asymmetry of secondary leptons w.r.t. the production plane of
$t$ quarks. The other two polarizations are more complicated asymmetries.
$\eta_3$ is the longitudinal polarization $P_\parallel$, $\eta_1$ is
the transverse polarization in the production plane of the $t$ quark
$P_\bot$ and $\eta_2$ is the transverse polarization perpendicular to the
production plane $P_N$. $\eta_2$ is odd under naive time reversal and hence is
sensitive to the absorptive part of the production matrix element. This can
arise either due to some new physics or simply due to the QCD corrections to
the production process. The other two degrees of polarization appear due to 
parity-violating interactions or simply due to the polarization of
particles in the initial state.

In order to calculate the top polarization directly in the lab frame using
Eq.~(\ref{polgen}), we need to measure the momentum of the lepton along
with $(\beta_t,\cos\theta_t)$ to evaluate the $s_i^\mu$. The easiest of 
the three
is the measurement of $\eta_2$ which requires the reconstruction of only the
$t$-production plane. The next is $\eta_1$ which requires $\cos\theta_t$ 
in addition to the production plane, and the most difficult is $\eta_3$ 
which requires
reconstruction of the full $t$-quark momentum, i.e. $\beta_t$ and
$\cos\theta_t$. A discussion of $t$-quark momentum reconstruction is 
given in Appendix~\ref{appA}.
This method of measuring polarization does not depend upon the
process of top production and hence can be applied to all processes and in
any frame of reference. The freedom to choose any frame of reference allows us
to consider a hadron collider or a photon collider without any additional
difficulty. The most important point here is that this measurement is not
contaminated by the anomalous decay of the $t$ quark and hence is a {\em true}
probe of its polarization. The effect of anomalous couplings or 
radiative corrections shows up in the energy distribution of the decay leptons.

\section{Energy distribution of secondary leptons}
From Eq.~(\ref{dsig2n}) it is clear that the energy distribution of the lepton
depends only on $F(E_\ell)$. The only way the effect of the production
process enters the $E_\ell$ distribution in the lab frame is through the boost
parameters $E_t$ and $\cos\theta_{t\ell}$. Since the $\cos\theta_{t\ell}$
distribution depends on the
polarization of the decaying $t$ quark, the $E_\ell$ distribution shows
sensitivity to the $t$ polarization. However, in the rest frame the $E_\ell^0$
distribution is completely independent of the $t$-production process and its
polarization. The only dependence in the rest frame is on the anomalous
couplings and can be used to measure them. In the following we will
demonstrate this feature of the $E_\ell$ distribution by considering
$t \bar t$ production at a photon linear collider (PLC).

\begin{figure}[t]
\centerline{\epsfig{file=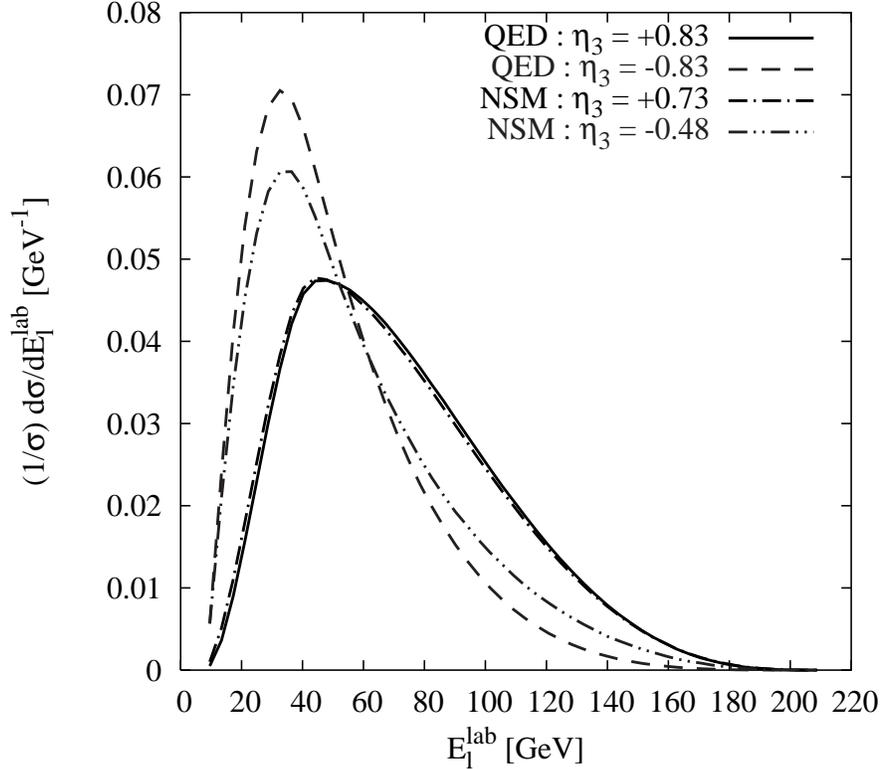,width=12.0cm}}
\caption{\label{fig:Elpol}The distribution in the energy of the lepton in the
lab frame with $P^+$ (positively polarized) and $P^-$ (negatively polarized)
initial states both for the pure QED contribution and 
for $t\bar t$ production
including the contribution of a non-standard (NSM) Higgs boson.}
\end{figure}

For simplicity, we consider a massless $b$ quark and a narrow-width approximation
for the $W$ boson. With the former assumption only $f_{2R}$ contributes to the
$E_\ell$ distribution.
We consider the production process $\gamma\gamma\to t\bar t$ with/without 
contribution from $\gamma\gamma \to \phi \to t \bar t$~\cite{phphtt2}, where
$\phi$ is a non-standard model (NSM) Higgs boson
of mass 475 GeV, width 2.5  GeV. Further, we take for the top couplings
($S_t$, $P_t$) and the $\gamma$ couplings ($S_\gamma$, $P_\gamma$) of the
Higgs boson, defined in Ref.~\cite{phphtt2}, the following arbitrarily
chosen values~:
$S_t =0.2, \ P_t=0.4, \ S_\gamma = 4.0 + i \ 0.5 \ \mbox{and} \ P_\gamma =
1.25 + i \ 2.0.$ We then  study the change in polarization of the $t$ quark
and its effect on the $E_\ell$ distribution.  We use the ideal photon
spectrum of~\cite{phspec} and calculate various kinematical distributions for
initial-state polarizations 
$$ P^+ \ \equiv \ \lambda_{e^-} = \lambda_{e^+} = +, \ \lambda_1=\lambda_2 = -
\hspace{0.5cm} \mbox{and} \hspace{0.5cm}
P^- \ \equiv \ \lambda_{e^-} = \lambda_{e^+} = -, \ \lambda_1=\lambda_2
= +.$$
For a PLC running at 600 GeV, the 
QED prediction for polarization is $+0.83$ with a
$P^+$ initial state and $-0.83$ with a $P^-$ initial state. The polarization in the presence
of a non-standard Higgs boson is $+0.73$ and $-0.48$ for $P^+$ and $P^-$ initial
states respectively. For the two choices of polarized initial states 
the $E_\ell$
distribution is shown in Fig.~\ref{fig:Elpol} for both QED and (NSM Higgs +
QED), where the latter is denoted by ``NSM''. Here $f_{2R}=0$ is assumed.
We see that the $E_\ell$ distribution
is peaked at lower values of $E_\ell$ when the $t$ quark is negatively polarized
and the peak of the  distribution is shifted to a higher value for positively
polarized $t$ quarks.
This can be understood as follows. In the rest frame of the $t$ quark the
angular distribution of leptons is $(1+\eta_3\cos\theta_\ell)$. Thus
for a positively polarized $t$ quark most of the decay leptons come in the forward
direction, i.e. the direction of the would-be momentum of the $t$ quark. Thus
a boost from the rest frame to the lab frame increases the energies of these 
leptons. This explains the shifted position of the peak for a positively 
polarized $t$ quark.  Similarly, for negative polarization most of the decay 
leptons come out in the backward direction w.r.t. the lab momentum of the 
$t$ quark. This results in an opposite boost and hence a decrease in the 
energy of the leptons. In other words, it leads to
increase in lepton counts for lower energy. This explains the large peak in
$E_\ell$ distribution at lower $E_\ell$. Further, for the case of $P^-$ initial
state, there are large modifications in the values of $t$ polarization due to
the Higgs contribution as compared to the pure QED prediction. These large 
differences show
up in the dashed and dashed-double-dotted curves in Fig.~\ref{fig:Elpol}.
\begin{figure}[th]
\centerline{\epsfig{file=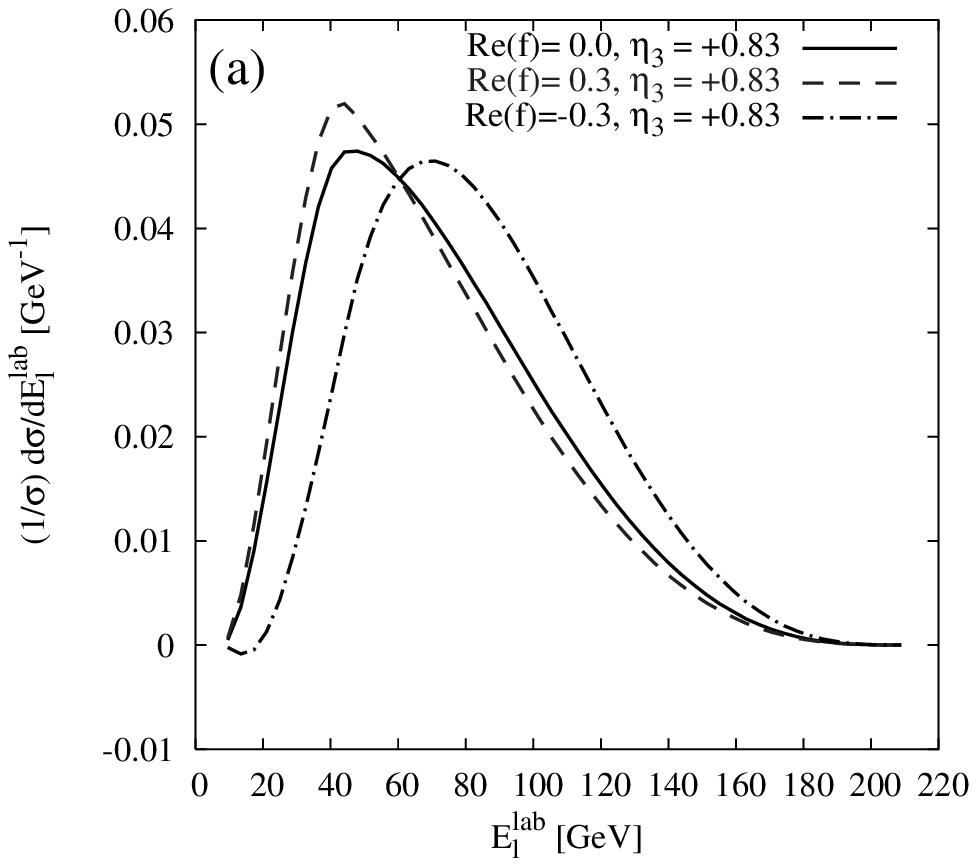,width=7.0cm}
\epsfig{file=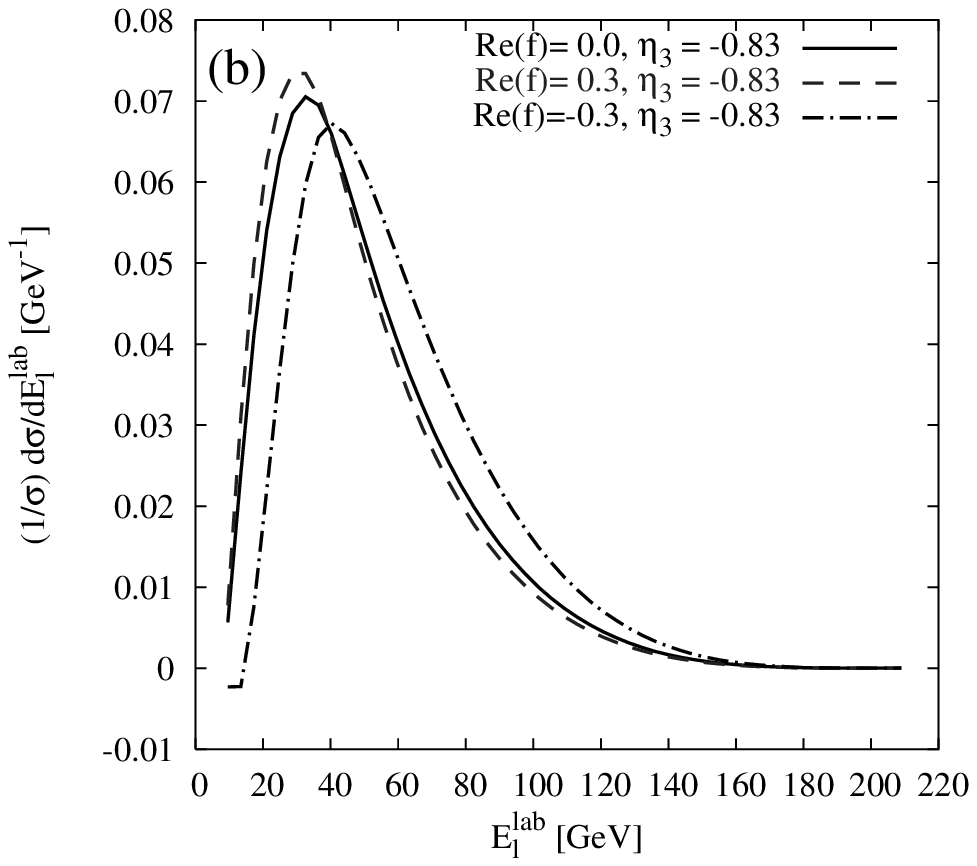,width=7.0cm}}
\caption{\label{fig:Ellab}The distribution in the energy of the lepton in the
lab frame within QED for different values of $\Re(f_{2R})$, with $P^+$ (left
panel) and $P^-$ (right panel) initial states.}
\end{figure}
The $E_\ell$ distribution is obviously dependent on the anomalous $tbW$
couplings and the dependence is shown in Fig.~\ref{fig:Ellab} for different
values of $f(=f_{2R})$ and initial states $P^+$ and $P^-$. Thus
we see that the $E_l$ distribution in the lab frame is affected by the
polarization of the decaying $t$ quarks as well as by the anomalous $tbW$
couplings. Thus $E_\ell$ distribution cannot be used as a definite signal for
either $t$ polarization or anomalous $tbW$ couplings in the lab frame due to
their intermingled effects.

\begin{figure}[th]
\centerline{\epsfig{file=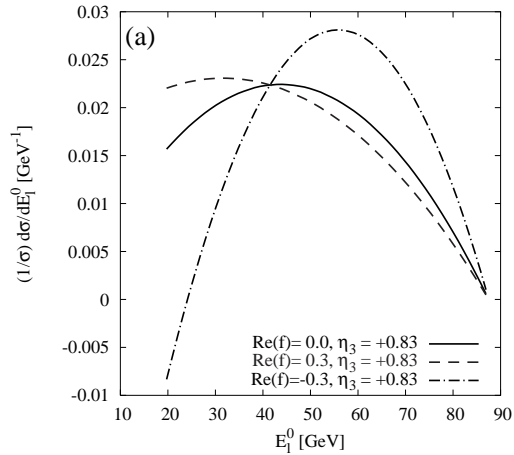,width=7.0cm}}
\caption{\label{fig:Elrest}The distribution in the energy of the lepton in the
rest frame within QED for different values of $\Re(f_{2R})$ with $P^+$ initial 
state.}
\end{figure}
In the rest frame of the $t$ quark, however, the angular and energy 
dependences are decoupled from each other. Hence the energy 
distribution is independent  of the $t$ polarization which in turn may depend 
on the production process. The $E_{\ell}^0$ distribution is given by
\begin{eqnarray}
\frac{1}{\sigma} \ \frac{d\sigma}{dE_\ell^0} = \frac{1}{\Gamma_t(\ell)} \ \left(
\frac{\alpha^2}{8 \ \sin^4\theta_W}\right) \ \frac{1}{\pi \ m_t} \ E_\ell^0
F(E_\ell^0) \ \left|\Delta_W(p_W^2)\right|^2 \ dp_W^2,
\end{eqnarray}
independent of the production process. Here $\Gamma_t(\ell)$ is the partial
decay width of the $t$ quark in the semi-leptonic channel. We note that
$\Gamma_t(\ell)$ depends upon anomalous $tbW$ couplings. For massless
$b$ quarks and on-shell $W$ bosons the $E_\ell$ distribution reduces to
\begin{eqnarray}
\frac{1}{\sigma} \ \frac{d\sigma}{dE_\ell^0} = \frac{1}{\Gamma_t(\ell)} \ \left(
\frac{\alpha^2}{8 \ \sin^4\theta_W}\right) \ \frac{1}{m_t \ m_W \ \Gamma_W} \
E_\ell^0 \ (m_t^2 - 2m_t \ E_\ell^0) \ \left( 1 + \Re(f_{2R}) \
\frac{m_W}{E_\ell^0}\right),
\end{eqnarray}
while the semi-leptonic partial decay width is given by
\[ \Gamma_t(\ell) = \left(
\frac{\alpha^2}{192 \ \sin^4\theta_W}\right) \ \frac{1}{m_t  m_W  \Gamma_W} \
\frac{(m_t^2 - m_W^2)^2}{m_t^2} \ \left(m_t^2+2m_W^2 + 6 \ \Re(f_{2R}) \
m_t m_W \right).
\]
Thus we see that the energy distribution of the decay lepton is completely
independent of the production process, the kinematical distribution of the $t$ quark
or its polarization. It depends only on $\Re(f_{2R})$ and  is shown in
Fig.~\ref{fig:Elrest}.  The distribution  shows a strong
dependence on $\Re(f_{2R})$. The crossing point
in the distribution is at $E_\ell^C= (m_t + 2m_W^2/m_t)/6 = 41.5$ GeV when
there is an accidental cancellation between $\Re(f_{2R})$-dependent terms in
$F(E_\ell)$ and $\Gamma_t(\ell)$. One can define an asymmetry about this
crossing point as
\begin{equation}
A_{E_\ell} = \frac{\sigma(E_\ell^0 > E_\ell^C) - \sigma(E_\ell^0 < E_\ell^C)}
{\sigma(E_\ell^0 > E_\ell^C) + \sigma(E_\ell^0 < E_\ell^C)} \ = \
\frac{13 m_t^2 - 22 m_W^2 -18 \Re(f_{2R}) \ m_tm_W}{27 \ (m_t^2+2 m_W^2 +
6 \Re(f_{2R}) \ m_t m_W)}.
\label{eq:asymf}
\end{equation}
This asymmetry is sensitive to $\Re(f_{2R})$, the anomalous $tbW$ coupling. 
If the four-momentum of the
decaying $t$ quark is fully reconstructed  the rest-frame lepton energy can
be computed as $E_\ell^0 = (p_t\cdot p_\ell)/m_t$ and the distribution shown in
Fig.~\ref{fig:Elrest} can be generated. Then using the asymmetry $A_{E_\ell}$
the value of $\Re(f_{2R})$ can be measured independent of any possible new
physics in the production process. This is another manifestation 
of the decoupling of the lepton angular distribution.

Thus, if $p_t$ can be fully reconstructed then the  spin-basis
vectors $s_i$ can be constructed. Using Eq.~(\ref{polgen}) one can then probe 
the polarization of the $t$ quark and  any new physics in 
the production process, independent of anomalous $tbW$ couplings using the 
angular distribution of the decay leptons. At the same time, using  the 
$E_\ell^0(=(p_t\cdot p_\ell)/m_t)$
distribution and $A_{E_\ell}(f)$, one can probe anomalous $tbW$ couplings
independent of the new physics in the production process of the $t$ quark.
It is interesting to note that the scalar product of $p_\ell$ with $p_t$ and
$s_i$
can probe effects of new physics in both production and decay 
processes of $t$ quarks. The quantity $(p_t\cdot p_\ell)$ is sensitive only to 
the new physics in the decay vertex independent of the production process or 
dynamics, while $(s_i\cdot p_\ell)$ are sensitive only to the production 
dynamics independent of anomalous contributions to the top decay vertex.

\section{\boldmath Simple and qualitative probes of $t$ polarization}
A completely decoupled analysis of possible new physics in production and decay
processes of the $t$ quark is possible. However, such an analysis necessarily
requires full reconstruction of the four-momentum of the $t$ quark. Full
reconstruction of $p_t$ is not always possible and it is useful to
to look for some easily measurable variables or distributions, which could 
probe $t$ polarization. The lab frame distribution of the
lepton energy shows sensitivity to the $t$ polarization, but it is contaminated
by possible presence of anomalous $tbW$ couplings. The lab frame lepton angular
distribution, on the other hand, is insensitive to the anomalous $tbW$ couplings
and can be used at least as a qualitative probe of $t$ polarization. For
demonstration purposes we again consider top-pair production at a PLC as 
in the last section.
\begin{figure}[th]
\centerline{\epsfig{file=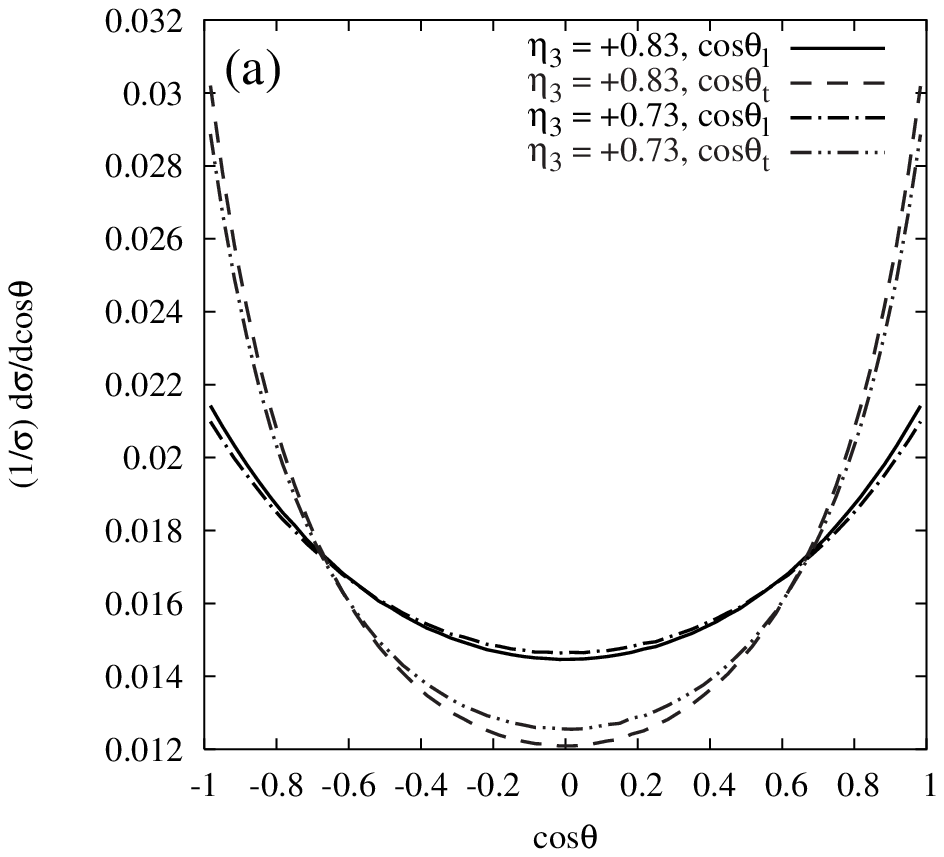,width=7.0cm}
\epsfig{file=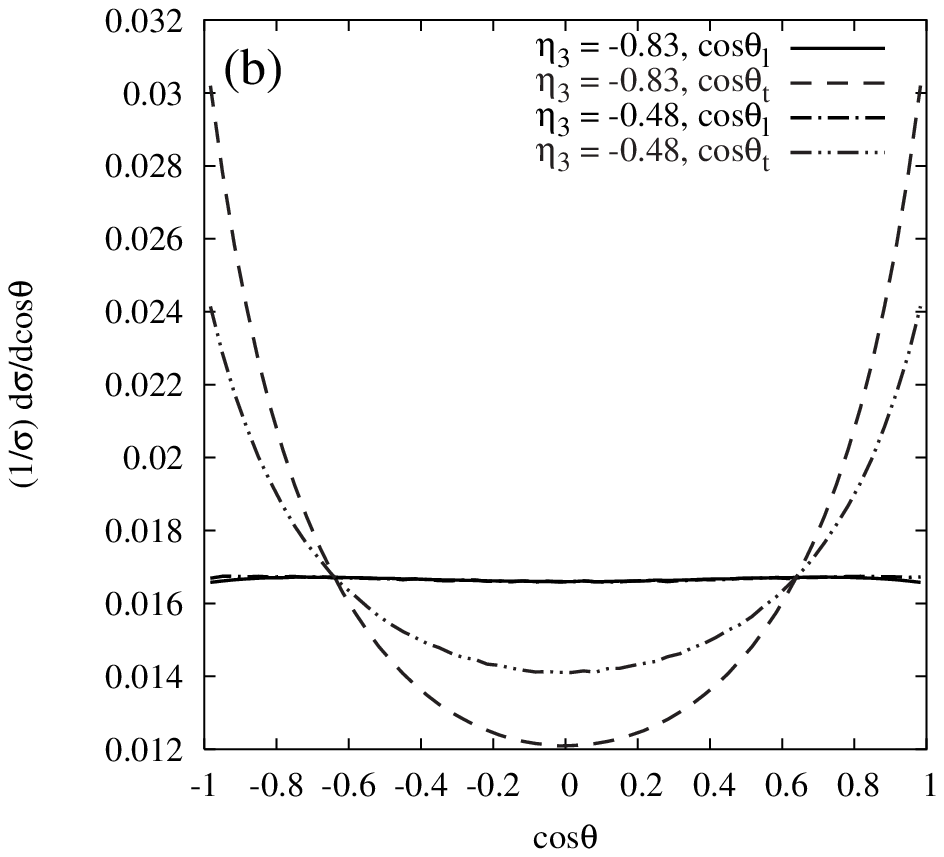,width=7.0cm}}
\caption{\label{fig:Cx}The distributions in the cosines of the angles of
the top and
the lepton, $\cos\theta_t$ and $\cos\theta_\ell$, respectively,
in the lab frame with $P^+$ (left panel) and $P^-$ (right panel) initial
states. For pure QED we have $\eta_3 = \pm 0.83$ while the presence of a Higgs
boson modifies it to $+0.73$ and $-0.48$ for $P^+$ and $P^-$ initial states,
respectively.}
\end{figure}
The $\cos\theta_\ell$ and $\cos\theta_t$ distributions with $P^+$ and
$P^-$ states are shown in Fig.~\ref{fig:Cx}. In Fig.~\ref{fig:Cx}(a), which is
drawn for the $P^+$ initial state, we see that the lepton distribution follows the
distribution of the $t$ quark in the lab frame up to some kinematical smearing.
On the other hand, for the $P^-$ initial state, the lepton distribution is flat,
i.e., it is completely smeared out. This is the effect of the polarization of the
$t$ quark, which is different in the two cases. For the pure QED case, the
distribution of the $t$ quark is exactly the same (the dashed line in both
Fig.~\ref{fig:Cx}(a) and (b)), while the polarization is $+0.83$ in the 
first case
and $-0.83$ in the second. Since positively polarized $t$ quarks have
leptons focused in the forward direction and  negatively polarized $t$ quarks
in the backward direction, the corresponding lepton distribution (solid line)
is quite different for the two cases in the lab frame. Any change in the $t$-quark
angular distribution, such as caused by the NSM Higgs boson
(dashed-double-dotted line), can also change the lepton polar distribution.
Secondly, one needs to know the $\cos\theta_t$ distribution, i.e., the
production  process, before hand in order
to estimate its polarization based on $\cos\theta_\ell$ distribution. Thus the
lepton polar distribution in the lab frame captures the effect of
$t$ polarization only in a qualitative way as in the case of the
$E_\ell$ distribution. One
advantage that the $\cos\theta_\ell$ distribution has is that it is insensitive to anomalous
$tbW$ couplings and depends only upon the dynamics of the production process.

\begin{figure}[th]
\centerline{\epsfig{file=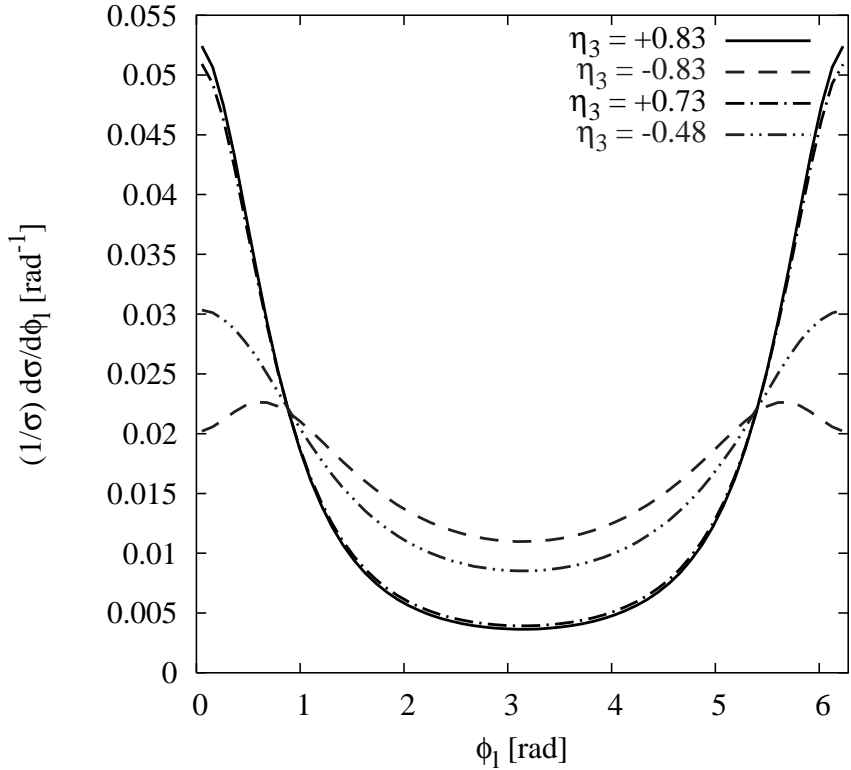,width=12.0cm}}
\caption{\label{fig:Ph}The distribution in the azimuthal angle of the lepton,
as defined in the text, in the lab frame with
$P^+$ and $P^-$ initial states for $t$ quarks with different polarizations.}
\end{figure}
The azimuthal distribution of the secondary leptons w.r.t. the production plane
of the $t$ quark also captures the effect of $t$ polarization in a qualitative
way. The skewness of the azimuthal distribution is related to $\eta_2$, 
the net
transverse polarization of the decaying $t$ quark perpendicular to the
production plane. $\eta_1$ and $\eta_3$ are degrees of polarization in the
production plane and lead to symmetric distribution about the production plane.
In the present case of $t\bar t$
production at a PLC through a Higgs boson, the net transverse polarization is
zero, i.e., $\eta_1=\eta_2=0$. Thus, the $\phi_\ell$ distribution is symmetric
about the $t$-production plane and shows sensitivity to $\eta_3$, the
longitudinal polarization, as shown in Fig.~\ref{fig:Ph}. We see that for
a positively polarized $t$ quark the $\phi_\ell$ distribution is peaked near
$\phi_\ell = 0 (2\pi)$ and the height of the peak decreases as the polarization
changes from $+0.83$ to $-0.83$. This again is related to the
$(1+\eta_3\cos\theta_\ell)$ distribution of the decay lepton in the rest frame
of the $t$ quark, which upon boosting experiences relativistic focusing and 
leads to a
larger peak for positive polarization and a smaller peak or suppression for negative
polarization. Unlike $E_\ell$ or $\cos\theta_\ell$ distributions, the
$\phi_\ell$ distribution can be used to quantify the $t$ polarization. The
up-down asymmetry is related to $\eta_2$ as shown in Eq.~(\ref{polgen}).
The peak height and the fractional area of the distribution near the peak
are qualitative measures of ``in-plane'' polarization of $t$ quarks. 

Thus, in conclusion, the angular
distribution of decay leptons itself provides a qualitative probe of the
$t$ polarization in the lab frame of the $\gamma\gamma$ collider. Similar
trends are expected for other colliders such as the LHC and LC and also 
for a general $2\to n$ process of $t$-quark production.

\section{\boldmath The $b$-quark angular distribution}
Even though the lepton angular and energy distributions offer a rather neat way
of probing $t$ polarization, this probe suffers from the rather low leptonic
branching ratio of the $W$ and the consequent small number of events that can
be used for the purpose. Indeed, this situation may be improved upon by using
the $b$-quark angular distribution. In this section we outline how this may be
done.

We consider a generic process of $t$-quark
production followed by $t\to b W$. The full differential cross-section is given
by
\begin{eqnarray}
d\sigma &=& \sum_{\lambda,\lambda'} \left[
\frac{(2\pi)^4}{2I} \rho(\lambda,\lambda')
\delta^4\bigg(k_A+k_B-p_t-\bigg(\sum_i^{n-1}p_i\bigg) \bigg)
\frac{d^3p_{t}}{2E_{t}(2\pi)^3} \ \prod_i^{n-1}
\frac{d^3p_{i}}{2E_{i}(2\pi)^3} \right]\nonumber \\
&\times&\left(\frac{1}{\Gamma_t}\right) \ \left[
\frac{(2\pi)^4}{2m_t} \Gamma(\lambda,\lambda')\delta^4(p_t-p_b-p_\nu-p_\ell)
\frac{d^3p_{b}}{2E_{b}(2\pi)^3} \ \frac{d^3p_{W}}{2E_{W}(2\pi)^3}
\right],
\label{facsigb}
\end{eqnarray}
for a narrow $t$ quark, similar to Eq.~(\ref{facsign}). The above equation assumes an
on-shell $W$ boson. The expression for the decay density matrix for $t\to bW$,
assuming a massless  $b$ quark, is obtained as
\begin{eqnarray}
\Gamma(\pm,\pm)&=& \frac{g^2 \ m_t^2}{2}\left[{\cal C}_1 \pm {\cal C}_2 \
\cos\theta_b\right] \nonumber \\
\Gamma(\pm,\mp)&=&\frac{g^2 \ m_t^2}{2} \ \left[ {\cal C}_2 \ \sin\theta_b \
e^{\pm\phi_b} \right],
\end{eqnarray}
where
\begin{eqnarray}
{\cal C}_1 &=& \left(\frac{1}{2} + \frac{m_t^2}{2 \ m_W^2} - \frac{m_W^2}{m_t^2}
\right) + 3\Re(f_{2R}) \left(\frac{m_t}{m_W} - \frac{m_W}{m_t} \right),
\nonumber\\
{\cal C}_2 &=& \left(\frac{3}{2} - \frac{m_t^2}{2 \ m_W^2} - \frac{m_W^2}{m_t^2}
\right) + \Re(f_{2R}) \left(\frac{m_t}{m_W} - \frac{m_W}{m_t} \right).
\end{eqnarray}
Thus, in the rest frame of the decaying the $t$ quark, the angular distribution of
the $b$ quark, similar to Eq.~(\ref{dsigrest}), is given by
\begin{eqnarray}
\frac{d\sigma}{d\cos\theta_b \ d\phi_b} = \frac{\sigma_{tot}}{4\pi} \ \left[
1 + \frac{{\cal C}_2}{{\cal C}_1} \ \left(\eta_3 \ \cos\theta_b + \eta_1 \
\sin\theta_b\cos\phi_b+ \eta_2 \ \sin\theta_b\sin\phi_b \right) \right].
\label{dsigrestb}
\end{eqnarray}
Hence the expressions for the $t$ polarization, similar to Eq.(\ref{polgen}),
can be written as
\begin{eqnarray}
\frac{\eta_3}{2} \ \frac{{\cal C}_2}{{\cal C}_1}&=&\frac{\sigma(p_b.s_3 < 0)
- \sigma(p_b.s_3 > 0)}{\sigma_{tot}} \nonumber\\
\frac{\eta_2}{2}\frac{{\cal C}_2}{{\cal C}_1}&=&\frac{\sigma(p_b.s_2 < 0)
- \sigma(p_b.s_2 > 0)}{\sigma_{tot}} \nonumber\\
\frac{\eta_1}{2}\frac{{\cal C}_2}{{\cal C}_1}&=&\frac{\sigma(p_b.s_1 < 0)
- \sigma(p_b.s_1 > 0)}{\sigma_{tot}}.
\label{polgenb}
\end{eqnarray}
Thus, for $b$-quark distributions the anomalous $tbW$ couplings enters through
factors ${\cal C}_1$ and ${\cal C}_2$, or rather,  through their ratio which
is given by
\begin{eqnarray}
\frac{{\cal C}_2}{{\cal C}_1}&=&-\left(\frac{m_t^2-2 \ m_W^2}{m_t^2+2 \ m_W^2}
\right) + \Re(f_{2R}) \ \left(\frac{8 \ m_t m_w (m_t^2-m_W^2)}{(m_t^2+2m_W^2)^2}
\right)\nonumber\\
&\equiv& \alpha_b^0 + \Re(f_{2R}) \ \alpha_b^1.
\end{eqnarray}
Here we have retained terms only up to linear order in $f_{2R}$. 
For $m_t=175$~GeV and $m_W=80.41$~GeV we have
\[ \alpha_b^0 = -0.406 , \hspace{1.0cm}\alpha_b^1= 1.43 \ .\]
In other words, the sensitivity of the $b$-quark distribution to the
$t$ polarization is less than that of the lepton distribution, the
analyzing powers being in the ratio
$[-0.406+ 1.43 \ \Re(f_{2R})~:~1.00]$. Thus the gain due to the larger
statistics is offset by the low sensitivity, and overall there is not much
gain. However if we consider $b$-quark angular asymmetries, Eq.~(\ref{polgenb}),
in association with the corresponding lepton angular asymmetries,
Eq.~(\ref{polgen}), we have
\begin{equation}
{\rm Br}(t\to b\ell\nu_\ell)\frac{ \left[\sigma(p_b.s_i < 0) - \sigma(p_b.s_i
> 0)\right]} {\sigma(p_\ell.s_i < 0) - \sigma(p_\ell.s_i > 0)} = \frac{{\cal
C}_2}{{\cal C}_1} = \alpha_b^0 + \Re(f_{2R}) \ \alpha_b^1.
\label{aratio}
\end{equation}
The ratio of the $b$-quark asymmetry to the leptonic asymmetry depends on
the anomalous $tbW$ coupling linearly and can be used to measure $\Re(f_{2R})$.
However, such a measurement is possible only if the $t$ polarization is large.
Considering only the semi-leptonic decay channel of the $t$ quark, the expected
limit on $\Re(f_{2R})$ is given by
\begin{equation}
|\Re(f_{2R})| \le \frac{f}{|\eta_i| \ \sqrt{{\cal L} \ \sigma}} \frac{\sqrt{1+
(\alpha_b^0)^2}}{\alpha_b^1} \ .
\label{limref}
\end{equation}
Here, ${\cal L}$ is the integrated luminosity, $\sigma$ is the total rate of
$t$-quark production and its semi-leptonic decay, $f$ is the degree of
statistical significance and $\eta_i$ is the average polarization of the
decaying $t$ quark. We note that the limit in Eq.(\ref{limref}) is independent
of the production mechanism of the $t$ quark but depends upon the average
polarization of the $t$ quark. The $t$-quark pair-production rate at an $e^+e^-$
collider is large and for polarized electron and positron beams the $t$ quark
is highly polarized. Hence it is the best and the cleanest place to measure
$\Re(f_{2R})$. Alternatively, one can undertake measurements 
at LHC where the $t\bar t$ production rate is very high $\sim 750$ pb\footnote{
Calculated using CompHEP.}.
QCD corrections may lead to $\eta_2\approx 10^{-3}$ in $q\bar q$ 
fusion~\cite{Bernreuther:1995cx} while possible new physics in the production
process may give a larger value of $\eta_2$.
Further, the measurement of $\eta_2$ requires re-construction of only the
production plane of the $t$ quark, which is possible at LHC\footnote{This
also requires the knowledge of the direction of initial quark momentum, 
which can be obtained only probabilistically.}. Thus one sees from 
Eq.(\ref{limref})  that for $i=2$, using the asymmetry of Eq.(\ref{aratio}) 
and assuming $|\eta_2|=0.01$, one may be
able to constrain $\Re(f_{2R})$ within $0.05$ at 95\% C.L. with about 
$8\times 10^7$ events for top quark. 
We emphasize that this estimate of number of events does not assume
anything about mechanism or kinematics of $t$ quark production.
For this analysis one has to look only at the semi-leptonic decay channel of
the $t$ quarks as that has rather small radiative corrections.

\section{Discussions and Conclusions}
The decoupling of decay-lepton angular distribution from anomalous
couplings in the $tbW$ vertex has been known for $e^+e^- \to t\bar t$~\cite{eett1,
eett2}, $\gamma\gamma \to t\bar t$~\cite{phphtt1, phphtt2} and also for a
general $2\to2$~\cite{decoup1, decoup2} process of $t$-quark pair production.
All these results have used the narrow-width approximation for $W$ bosons
and except for Ref.~\cite{decoup2} all of them assume a massless $b$ quark. The
vanishing mass  of the $b$ quark provides additional chiral symmetry and among the
anomalous couplings shown in Eq.~(\ref{V:tbW}) only $f_{2R}$ contributes. In the
present work we extended this decoupling theorem to a general $2\to n$ process
of $t$-quark production with a massive  $b$ quark (hence keeping all four
anomalous $tbW$ couplings)  and without using the narrow-width approximation for
$W$ bosons.  We analyzed the essential inputs for the decoupling and found that
the narrow-width approximation for $t$ quarks and smallness of $\{f_{1R}, \ f_{2L},
\ f_{2R}\}$ are the only two requisites for decoupling. This decoupling can also
be extended to the hadronic decay of $W$ bosons where the role of $\ell$ is
taken up by the down-type quark, i.e. $T_3=-1/2$ fermion in the $SU(2)$ doublet.
The charge measurement of the light-quark jets is the only technical barrier in
using this channel. We argue that within the narrow-width approximation, the
decoupling of the lepton angular distribution remains valid after radiative
corrections, while noting that the decoupling of the angular distribution
of the down-type quark receives about 7\% correction from QCD
contributions.

The polarization of the $t$ quark reflects in the kinematic
distribution of its decay products. We use the decoupled lepton angular
distribution to construct specific asymmetries (Eq.~(\ref{polgen})) to measure 
the $t$ polarization. These asymmetries,  robust against radiative corrections
since they are constructed after taking ratios, are insensitive to the 
anomalous $tbW$ coupling.  A full reconstruction of the four-momentum of the 
$t$ quark is necessary to construct these asymmetries experimentally, which 
may be possible only at the ILC.  At the LHC or the PLC one can construct the 
$t$-production plane and hence $\eta_2$ can
be constructed, which is sensitive to the absorptive part of the production
amplitude. The energy distribution of the decay lepton shows sensitivity to 
new physics in the decay process, i.e. the $tbW$ vertex. In the lab frame it
receives contribution both from the polarization of the $t$ quark and possible
anomalous $tbW$ couplings. However, in the rest frame of the $t$ quark it is
sensitive only to the anomalous $tbW$ couplings and is independent of the 
type of production process as well as
any possible new physics in the production process. Thus given full
reconstruction of the $t$ quark four-momentum, possible new physics in production
and decay processes of the 
$t$ quark can be studied independent of each other using
angular and energy distributions of secondary leptons, respectively.

We also studied the effect of $t$ polarization on the lepton angular 
distribution in the lab frame. Such an analysis is useful where
momentum reconstruction is not possible. We see that the $\cos\theta_\ell$
and $\phi_\ell$ distributions in the lab frame of a
$\gamma\gamma$ collider are sensitive to the polarization of the decaying top,
at least in a qualitative way. A quantitative estimate is possible for  
$\eta_2$, which can be obtained using the up-down asymmetry of decay leptons 
w.r.t. the $t$-production plane.

To summarize, we have demonstrated that the lepton distribution can be used to 
probe new physics contributions in the production and the decay processes of 
the $t$ quark, separately, independent of each other.
The lepton angular distribution is shown to be completely insensitive to
any anomalous $tbW$ coupling assuming  the $t$ quark to be on-shell and 
anomalous $tbW$ couplings to be small. The energy distribution in the rest
frame of the $t$ quark, on the other hand, was found to be sensitive only to the
anomalous $tbW$ couplings. We construct specific asymmetries involving the 
lepton
angular distribution w.r.t. the top momentum, to measure its polarization in a
generic process of $t$-quark production. The qualitative features of the
lab-frame angular distribution of the decay leptons have been shown to be 
sensitive to the polarization of the decaying $t$ quark.
\vskip .5cm

{\Large \bf Acknowledgments}\\[0.5cm]
\nopagebreak
This work was partially supported by Department of Science and Technology 
project SP/S2/K-01/2000-II and Indo French Centre for Promotion of Advanced
Research Project 3004-2. We also thank the funding agency Board for Research in
Nuclear Sciences  and the organizers of the 8th Workshop on High Energy Physics
Phenomenology (WHEPP8), held at the Indian Institute of Technology, Mumbai, 
January 5-16, 2004, where this work was initiated.

\appendix
\boldmath
\section{Momentum reconstruction for the $t$ quark}
\label{appA}
In a generic reaction of $t$-quark production and subsequent decay of unstable
particles, (partial) reconstruction of $p_t$ is possible if there is only
one or no missing neutrino in the final debris. For the semi-leptonic decay of
$t$ quarks there is one neutrino, thus we demand that all other particles in the
production part are observable or decay into observable particles. The cases of
different colliders are discussed.

{\bf Fixed $\sqrt{s}$ collider~:} The $e^+e^-$ linear collider is an example of
a fixed $\sqrt{s}$ collider. At these colliders the net three-momentum of the 
various particles should add up to zero. Since we have only one neutrino, its 
three-momentum can be determined and hence $p_t$ can be fully reconstructed. 
Thus the study of polarization of the $t$ quark at an $e^+e^-$ collider 
is possible through the analysis of the decay-lepton angular distributions.

{\bf Variable $\sqrt{s}$ colliders~:} Hadron colliders and photon colliders are
variable $\sqrt{s}$ colliders. The c.m. frame of colliding partons moves with 
an unknown momentum along the beam axis of the collider. Thus the three-momenta
of the final state particles should add up to $\vec{P}_{cm}$ which is parallel 
to the $z$ axis. This implies that the transverse momentum of all particles
should add up to zero. This immediately gives the transverse momentum of neutrino
and hence transverse momentum of the $t$ quark. This defines the production
plane of the $t$ quark and one can construct $\eta_2$, the transverse
polarization of the $t$ quark normal to the production plane. Since construction of
$\eta_2$ is possible at LHC and since the top is produced through QCD
interaction at LHC, one can study the absorptive part of the QCD corrections to
the production process via lepton distributions.

\end{document}